\documentclass[aip,amsmath,amssymb,reprint,]{revtex4-1}

\usepackage{graphicx}
\usepackage{dcolumn}
\usepackage{bm}

\usepackage[utf8]{inputenc}
\usepackage[T1]{fontenc}
\usepackage{mathptmx}
\usepackage{etoolbox}
\usepackage{float}
\usepackage{placeins}

\makeatletter
\def\@email#1#2{%
 \endgroup
 \patchcmd{\titleblock@produce}
  {\frontmatter@RRAPformat}
  {\frontmatter@RRAPformat{\produce@RRAP{*#1\href{mailto:#2}{#2}}}\frontmatter@RRAPformat}
  {}{}
}%
\makeatother
\begin{document}

\preprint{AIP/123-QED}

\title[Jacobian determinant as a deformation field in static billiards]{Jacobian determinant as a deformation field in static billiards}

\author{Anne Kétri P. da Fonseca}
\affiliation{Departamento de Física, UNESP - Universidade Estadual Paulista, Rio Claro, SP, Brazil}
\email{anne.ketri@unesp.br}

\author{André L. P. Livorati}
\affiliation{Departamento de Física, UNESP - Universidade Estadual Paulista, Rio Claro, SP, Brazil}
\email{anne.ketri@unesp.br}

\author{Rene O. Medrano-T}
\affiliation{Departamento de Física, Instituto de Ciências Ambientais, Químicas e Farmacêuticas, UNIFESP, Diadema, SP, Brazil}

\author{Diego F. M. Oliveira}
\affiliation{School of Electrical Engineering and Computer Science, University of North Dakota, USA}

\author{Edson D.\ Leonel}
\affiliation{Departamento de Física, UNESP - Universidade Estadual Paulista, Rio Claro, SP, Brazil}

\date{\today}

\begin{abstract}
We develop a deformation-based framework for analyzing static billiard systems through the Jacobian determinant computed in noncanonical angular coordinates. Although these systems are conservative, the determinant is not identically equal to unity, generating structured domains of local phase–space expansion ($\det J>1$) and contraction ($\det J<1$). We show numerically that these domains balance globally, providing a geometric manifestation of area preservation in noncanonical variables. The curves defined by $\det J=1$ act as deformation boundaries that intersect unstable periodic points and correlate with invariant manifolds. We prove analytically that period-two orbits restore exact unit determinant under composition, while higher-period orbits exhibit angular modulation consistent with reversibility. The Jacobian determinant thus reveals an additional geometric layer in phase–space organization and offers a complementary perspective on conservative billiard dynamics.
\end{abstract}

\maketitle

\begin{quotation}
Area preservation is a defining feature of conservative billiard dynamics, yet its geometric expression depends crucially on the choice of variables. When the dynamics is written in noncanonical angular coordinates, the Jacobian determinant ceases to be identically equal to unity, despite the absence of dissipation. This apparent paradox generates structured domains of local expansion and contraction across phase space. We show that the boundaries of these domains, defined by $\det J=1$, intersect unstable periodic points and mirror invariant manifolds, revealing an intrinsic deformation structure underlying the dynamics. Although local stretching and compression occur, their global balance restores area preservation. Interpreting the Jacobian determinant as a deformation field uncovers an additional geometric layer in the organization of static billiards and complements the traditional manifold-based description of conservative systems.
\end{quotation}

\section{\label{Introduction} Introduction}
The topology of phase space is a foundational aspect in the study of dynamical systems \cite{briefhistory}. When discrete mappings are used to describe the temporal evolution from the $n$-th to the $(n+1)$-th state, the resulting sequence of states defines an orbit. Phase space is the set of all admissible states of the system, and each orbit corresponds to a trajectory within it, making it possible to visualize invariant structures and transport mechanisms that organize the dynamics \cite{lichtenberg}. The Jacobian matrix provides a natural tool to characterize the local deformation induced by the mapping. Assuming a function $F:\mathbb{R}^m \rightarrow \mathbb{R}^k$ from a $m$-dimensional to a $k$-dimensional Euclidean space, defined by $k$ component functions $F_1(x_1,...,x_m),...,F_p(x_1,...,x_k)$, the associated Jacobian matrix is
\begin{equation}
J=\begin{bmatrix} 
    \frac{\partial F_1}{\partial x_1} & \dots & \frac{\partial F_1}{\partial x_m} \\
    \vdots & \ddots &  \vdots\\
    \frac{\partial F_k}{\partial x_1} & \dots & \frac{\partial F_k}{\partial x_m}
    \end{bmatrix}.
\end{equation}
This $k\times m$ matrix collects the partial derivatives of $F$ and represents the linear transformation induced on the tangent space at a given point. For a two-dimensional map ($k=m=2$), the determinant of the Jacobian provides the local area expansion or contraction factor. A mapping is area-preserving when $\det J=1$. In two dimensions, this condition is equivalent to the symplectic condition \cite{lichtenberg}, which is satisfied by Hamiltonian flows written in canonical variables \cite{Arnold}. This property is particularly relevant since many dynamical systems can be formulated in terms of Hamiltonians of the form $H(x,p,t)={\mathcal{P}}^2/2m+ V(x,t)$, including billiard systems \cite{ref1,livoratifaixas,ref3,ref4}.

Billiards are mathematical models in which a particle, or a set of noninteracting particles, undergoes specular collisions with a rigid boundary \cite{chernov}. The boundary may be static or time-dependent. In the static case, the potential $V(x,t)=V_0(x) + V_1(x,t)$ satisfies $V_1(x,t)=0$, with $V_0=0$ inside the boundary and infinite elsewhere. The boundary shape determines the dynamical properties of the system, leading to integrable, chaotic, ergodic, or mixed behavior \cite{sinai}. The circular billiard is integrable, with boundary $R(\theta)=1$ and the mapping for the angles $\theta$ and $\alpha$ given by $\theta_{n+1}=\theta_n + \pi - 2\alpha_n$ and $\alpha_{n+1}=\alpha_n$ \cite{chernov}. Mixed phase space can arise when the circle is deformed. The elliptical-oval boundary can be written in polar coordinates as \cite{onthe}
\begin{equation}
R(\theta,p,q,e,\epsilon)=\frac{1-e^2}{1+e\cos(q\theta)}+\epsilon \cos (p\theta).
\label{front}
\end{equation}
Here $\epsilon$ and $e$ control the boundary deformation: $\epsilon$ is associated with the oval component, with $\epsilon=0$ reducing Eq.~(\ref{front}) to the elliptical boundary, whereas $e$ deforms the ellipse, recovering the oval boundary for $e=0$. The circular billiard is recovered for $\epsilon=e=0$. The parameters $p$ and $q$ introduce additional geometric modulations and must be integers; non-integer values would generate discontinuities in the boundary and allow particle escape. For $p = 1$ one obtains the oval billiard, while for $p > 1$ the so-called oval-like billiards, with analogous considerations for $q$. The oval billiard with $e=0$ and $\epsilon \neq 0$ is nonintegrable and exhibits islands of periodicity and invariant spanning curves that delineate distinct chaotic regions \cite{livrodenisspringer}. In contrast, the elliptical billiard with $\epsilon =0$ and $e \neq 0$ remains integrable, but a separatrix emerges, dividing rotator and librator orbits \cite{berry}.

In billiard systems, the canonical pair is usually $(\theta, \sin\alpha)$ \cite{chernov}. Therefore, when the dynamics is expressed in the variables $(\theta,\alpha)$, the mapping is not written in canonical coordinates and its Jacobian determinant is generally not equal to one. Values $\det J \neq 1$ then quantify local deformation of phase-space elements in these coordinates, even though the underlying static billiard dynamics is conservative. While the pointwise value of $\det J$ depends on the chosen coordinates, we show that its spatial organization across phase space is highly structured and correlates with dynamical landmarks. This motivates the central perspective adopted here: rather than focusing on the pointwise condition $\det J=1$, we investigate the balance between the regions of local expansion $(\det J > 1)$ and contraction $(\det J < 1)$. We also examine how the geometry and boundaries of these regions relate to stable and unstable fixed points, whose explicit determination is not trivial in this class of systems. In particular, we show that fixed points, especially of period two, are associated with an exact restoration of area preservation under the composed map, and that the trajectories linked to such points provide additional insight into the organizing mechanisms of billiard phase space.

This paper is organized as follows: Section \ref{Sec2} describes the model, the mapping and the expression for the Jacobian determinant for the static elliptical-oval billiard. Section \ref{Sec3} presents the results for the Jacobian determinant along different phase spaces, as well as the values obtained for ratio between the number of trajectories inside regions of expansion and contraction. In Section \ref{Sec4} we connect these results to identification of fixed points in the system, as well as the characterization of the determinant in such points. Section \ref{Sec5} presents the final remarks, summary and conclusions. The derivation of the results for $\det J_F^2=1$ and $\det J_F^n$ for $n>2$ is detailed in the Appendix.

\section{\label{Sec2} Model, mapping and Jacobian determinant}

\begin{figure}
\includegraphics[width=\columnwidth]{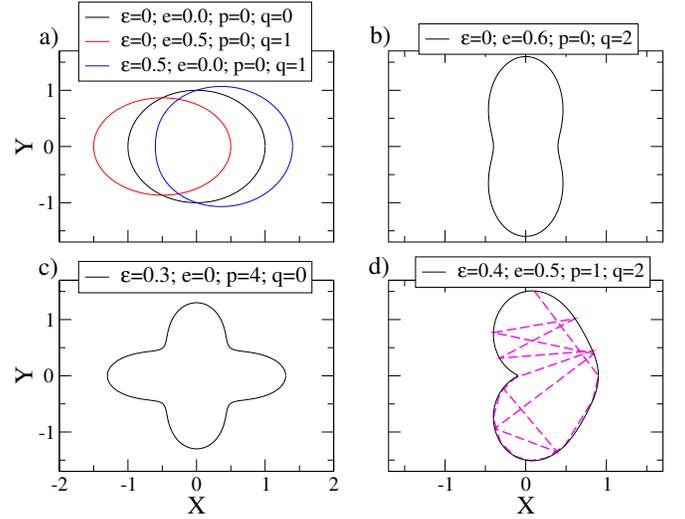}
\caption{\label{Fig1} Geometry of the boundary for the elliptical-oval billiard with different values of the control parameters $\epsilon$, $e$, $p$ and $q$. In \textit{d)} the trajectory of a single particle is shown for the first 20 collisions with $(\theta_0,\alpha_0)=(1.5,1.5)$.}
\end{figure}

As mentioned, the model consists of a particle, or a set of noninteracting particles, confined within a closed region and undergoing elastic and specular collisions with the boundary. The radius of the boundary, in polar coordinates, is given by Eq.~(\ref{front}). The control parameters $\epsilon$ and $e$ deform the circle: $e$ is associated with the elliptical shape, while $\epsilon$ introduces the oval component. Fig.~\ref{Fig1} shows the geometry of the boundary for different values of the control parameters $\epsilon$, $e$, $p$ and $q$. Fig.~\ref{Fig1} \textit{d)} also displays the trajectory of a single particle for the first 20 collisions with initial condition $(\theta_0,\alpha_0)=(1.5,1.5)$.

\begin{figure}
\includegraphics[width=\columnwidth]{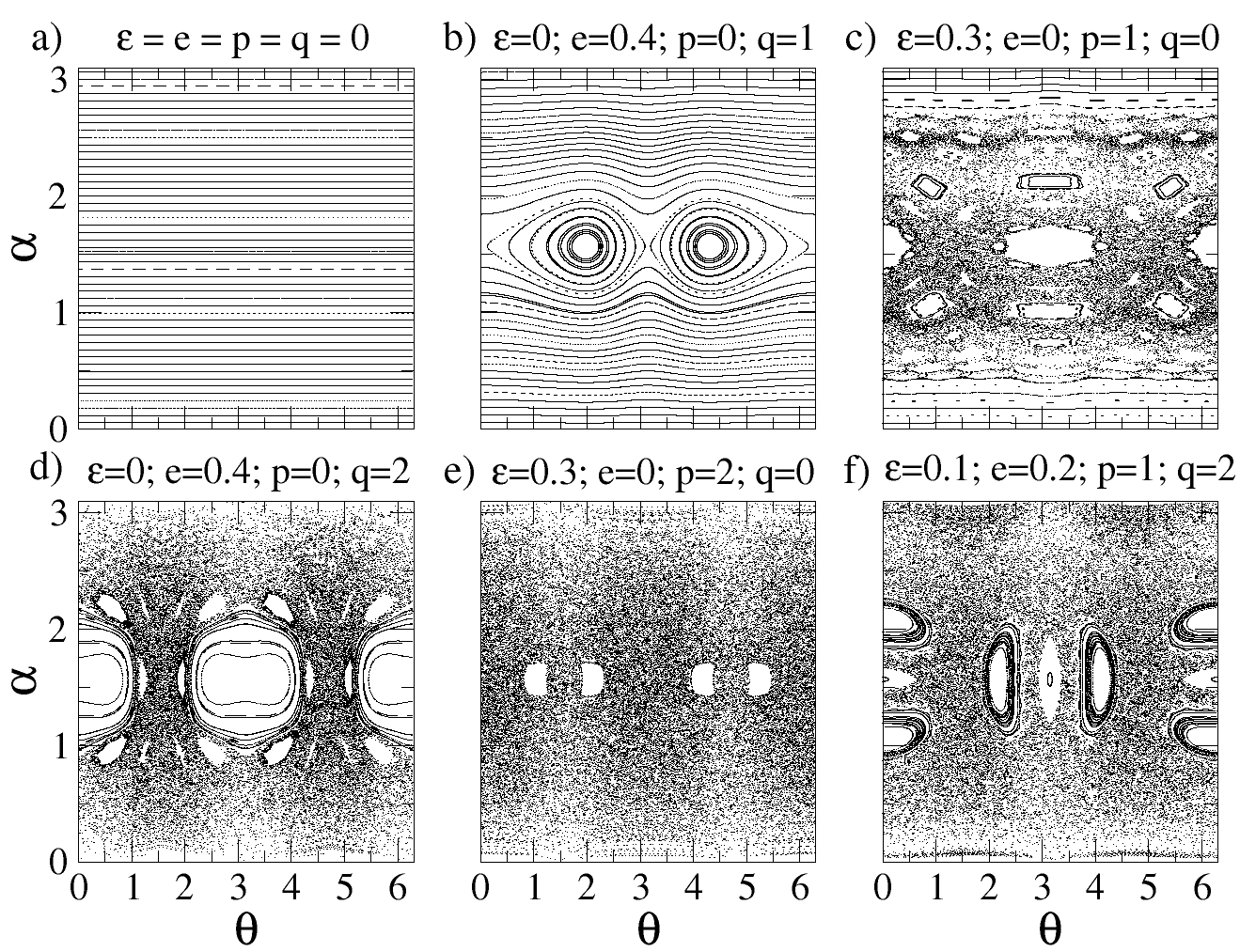}
\caption{\label{Fig2} Phase space for the elliptical-oval billiard considering different sets of control parameters $\epsilon$, $e$, $p$ and $q$.}
\end{figure}

Once the boundary is defined, the particle position can be written as
\begin{gather}
    X(\theta_n)=R(\theta_n)\cos(\theta_n), \\
    Y(\theta_n)=R(\theta_n)\sin(\theta_n).
\end{gather}
The dynamical variable $\theta_n$ denotes the angular position of the particle relative to the origin. The angle between the tangent and the boundary at this position is
\begin{equation}
    \phi_n=\arctan\left[\frac{Y'(\theta_n)}{X'(\theta_n)}\right],
\end{equation}
where $X'(\theta_n)$ and $Y'(\theta_n)$ denote the first derivatives of Eqs. (3) and (4) with respect to $\theta$. The trajectory of the particle between two successive collisions, assuming constant velocity, is then described by
\begin{equation}
    Y(\theta_{n+1})-Y(\theta_n)=\tan(\alpha_n+\phi_n)[X(\theta_{n+1})-X(\theta_n)],
\end{equation}
where $\alpha_n$ is the angle between the trajectory and the tangent vector to the boundary at $\theta_n$. Finally, the two-dimensional nonlinear mapping $T(\theta_n,\alpha_n)=(\theta_{n+1},\alpha_{n+1})$ is 
\begin{eqnarray}
   T:\begin{cases}  h(\theta_{n+1})=R(\theta_{n+1})\sin(\theta_{n+1})-Y(\theta_n) -\tan(\alpha_n+\phi_n)\\ \ \ \ \ \  \ \  \ \ \ \ \ \ \ \  \ \ \   \times [R(\theta_{n+1})\cos(\theta_{n+1})-X(\theta_n)]\\
    \alpha_{n+1}=\phi_{n+1}-(\alpha_n+\phi_n).\end{cases}
\label{map}
\end{eqnarray}
The value of $\theta_{n+1}$ is obtained numerically from the condition $h(\theta_{n+1})=0$, while the expression for $\alpha_{n+1}$ follows from geometrical considerations \cite{onthe}. In Fig.~\ref{Fig2} we present the phase space generated by iterating Eq.~(\ref{map}) for different sets of control parameters.

In Fig.~\ref{Fig2} \textit{a)} the fully integrable phase space of the circular billiard is shown, characterized by invariant straight lines and periodic orbits. Fig.~\ref{Fig2} \textit{b)} presents the phase space of the elliptical billiard, which remains integrable but exhibits a large double island separated by a separatrix and surrounded by invariant spanning curves. Fig.~\ref{Fig2} \textit{c)} shows the mixed phase space of the oval billiard, where chaotic regions coexist with regular structures such as periodic islands and invariant spanning curves. Fig.~\ref{Fig2} \textit{e)} illustrates a fully chaotic phase space after all regular structures have been destroyed. The critical value for the destruction of the last invariant curve is $\epsilon_c=\frac{1}{1+p^2}$, obtained from the variation of the boundary curvature \cite{critic}. Chaotic behavior is also observed in Fig.~\ref{Fig2} \textit{d)} and \textit{f)}. In the former, $q\neq 1$ leads to an elliptical-like billiard with characteristics similar to the oval case. In the elliptical-oval configuration, the combination of parameters generates multiple periodic structures distributed across phase space and embedded in a chaotic sea.

From the mapping in Eq.~(\ref{map}), the Jacobian matrix for $\theta_{n+1},\alpha_{n+1}$ is
\begin{equation}
    J=\begin{pmatrix}
\frac{\partial\theta_{n+1}}{\partial\theta_n} &  \frac{\partial\theta_{n+1}}{\partial\alpha_n} \\
\frac{\partial\alpha_{n+1}}{\partial\theta_n} & \frac{\partial\alpha_{n+1}}{\partial\alpha_n} 
\end{pmatrix}
\end{equation}
which leads to \cite{critic}
\begin{equation}
    \det J= -\frac{Y'(\theta_n) - X'(\theta_n)\tan(\phi_n + \alpha_n)}{Y'(\theta_{n+1}) - X'(\theta_{n+1})\tan(\alpha_n+\phi_n)}.
\label{detJ}
\end{equation}
As mentioned, this quantity is not identically equal to unity because $(\theta,\alpha)$ is not the canonical pair of variables for this system. Along phase space, different values of this pair lead to values of Eq.~(\ref{detJ}) associated with local deformation: expansion $(\det J > 1)$ or contraction $(\det J < 1)$. In what follows we use this determinant as a geometric observable on phase space, and investigate how its spatial organization correlates with the underlying dynamical structures.

\section{\label{Sec3} Visualization of Jacobian determinant properties}
\begin{figure}[h]
\includegraphics[width=\columnwidth]{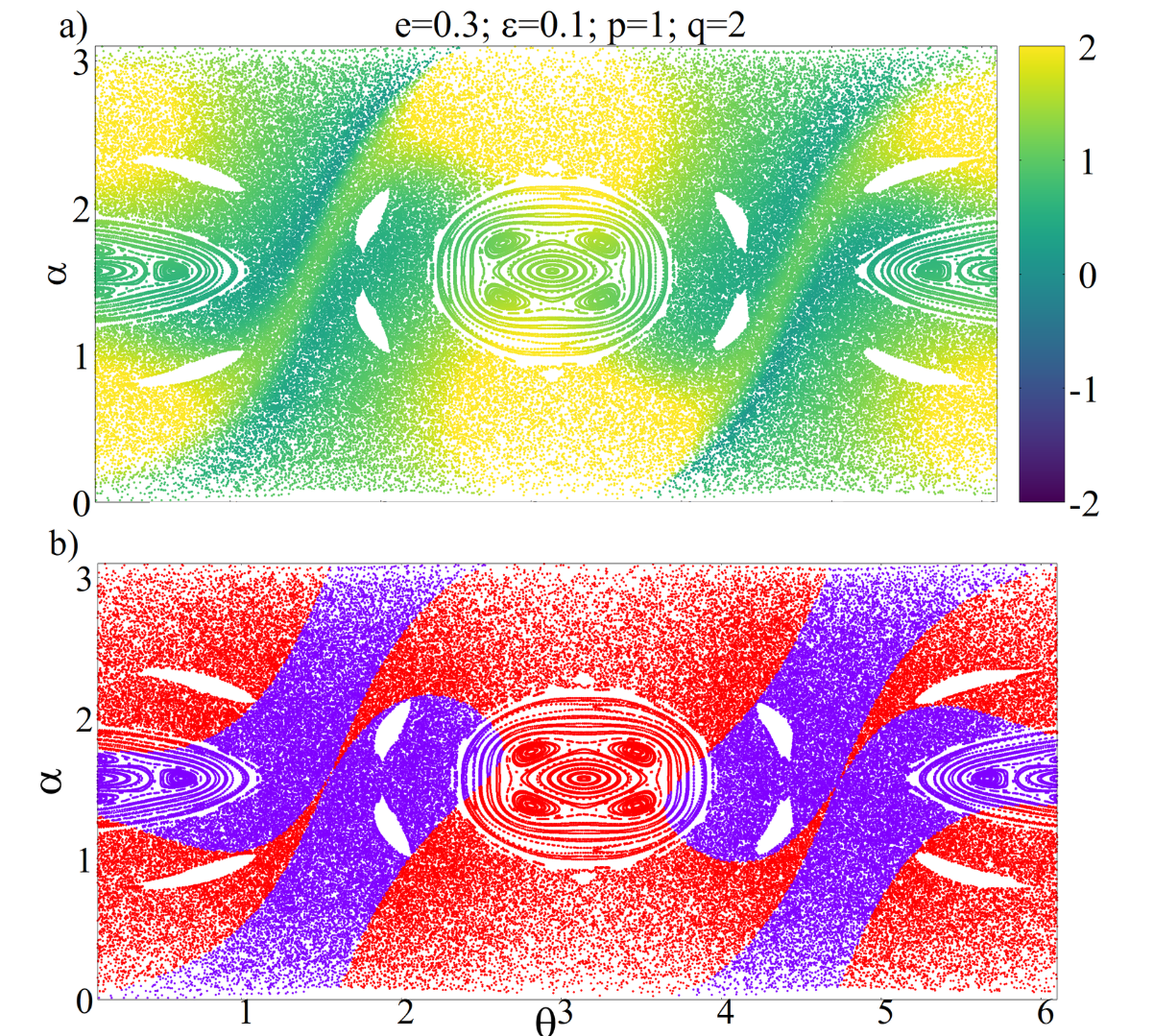}
\caption{\label{Fig3} Phase space and determinants of the Jacobian for the static elliptic-oval billiard with $e=0.3$,$\epsilon=0.1$,$p=1$ and $q=2$. The color scale represents the values of the determinant $\det J$, as given by Eq.~(\ref{detJ}). Panel (a) shows a continuous range from $-2$ to $2$, whereas panel (b) highlights only two regions: red for $\det J>1$ and blue for $\det J<1$.}
\end{figure}

Once the expression for the Jacobian determinant of the elliptical-oval billiard is obtained (see Eq.~(\ref{detJ})), we return to the phase space analysis. The same structures presented in Fig.~\ref{Fig2}, generated by iterating the mapping in Eq.~(\ref{map}), are now colored according to the value of the determinant. The results, together with the range of values used, are shown in Fig.~\ref{Fig3} \textit{a)} and reveal additional geometric organization across phase space.

\begin{figure*}
\includegraphics[width=2.0\columnwidth]{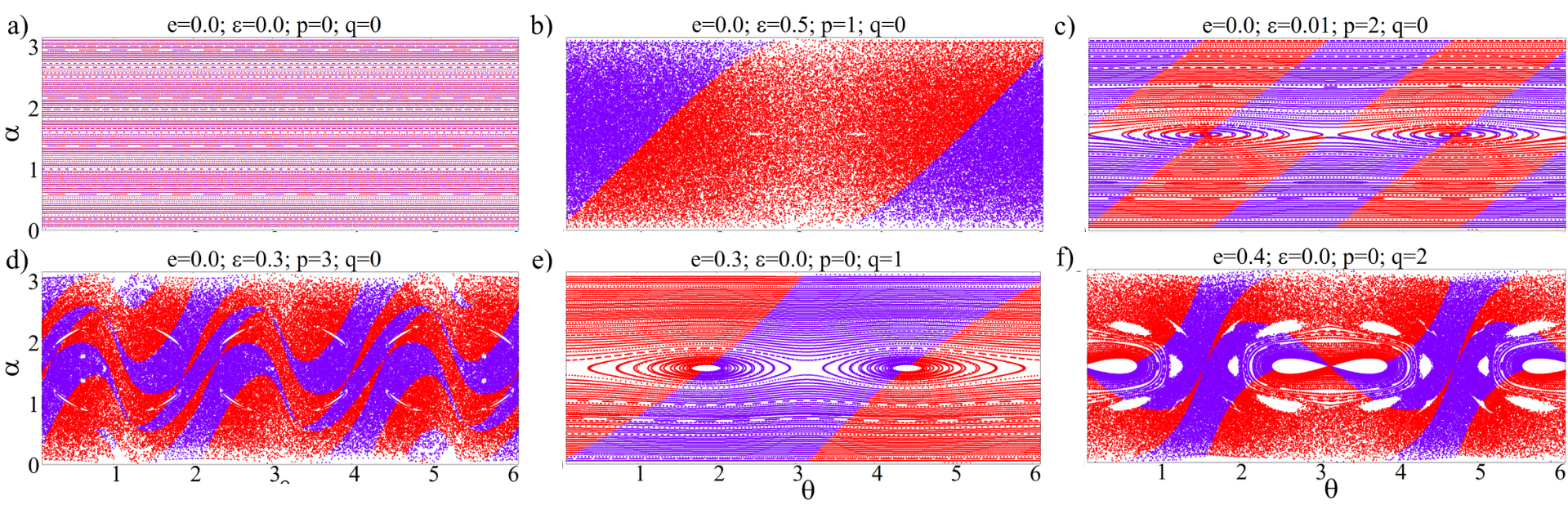}
\caption{\label{Fig4} Phase spaces and determinants of the Jacobian for the static elliptic-oval billiard with different values of $e$, $\epsilon$, $p$ and $q$. Determinant values below the unity are represented in blue, while those above unity are represented in red.}
\end{figure*}

To further clarify these structures, we classify determinant values into two regions: red for $\det J>1$ and blue for $\det J<1$ \cite{livoratifaixas}. The results, shown in Fig.~\ref{Fig3} \textit{b)}, make the expansion and contraction domains explicit. This procedure is then extended to different control parameters and geometries, as presented in Fig.~\ref{Fig4}.

By comparing the phase spaces, we observe that once we depart from the fully integrable circular billiard (Fig.~\ref{Fig4} \textit{a)}), the colored regions become clearly separated. For small values of $\epsilon$ and $e$, these domains appear as parallelogram-like structures periodically distributed throughout phase space. As the control parameters increase, the domains progressively deform and may merge, while their boundaries remain sharply defined. This behavior is observed for the elliptical, oval and elliptic-oval billiards. Comparing panels \textit{b)} and \textit{e)} of Fig.~\ref{Fig4}, one also identifies an inversion of the blue and red regions between the strictly oval and elliptical billiards, reflecting the distinct boundary deformations illustrated in Fig.~\ref{Fig1}.

The influence of the control parameters $p$ and $q$ can also be identified in Fig.~\ref{Fig4}. Increasing their values leads to a proliferation of the colored domains. By comparing panels \textit{b)}, \textit{c)} and \textit{d)} of Fig.~\ref{Fig4}, one observes the sequence $(2,8,18)$ for the number of regions. Further inspection shows that this pattern follows the rule $2p^2$, valid for integer values of $p>1$ and $e=q=0$. An analogous result holds in the elliptical case, with inversion of the blue and red regions, where the corresponding rule becomes $2q^2$ for integer values of $q>1$ and $\epsilon=p=0$.

\begin{figure}[h]
\includegraphics[width=0.9\columnwidth]{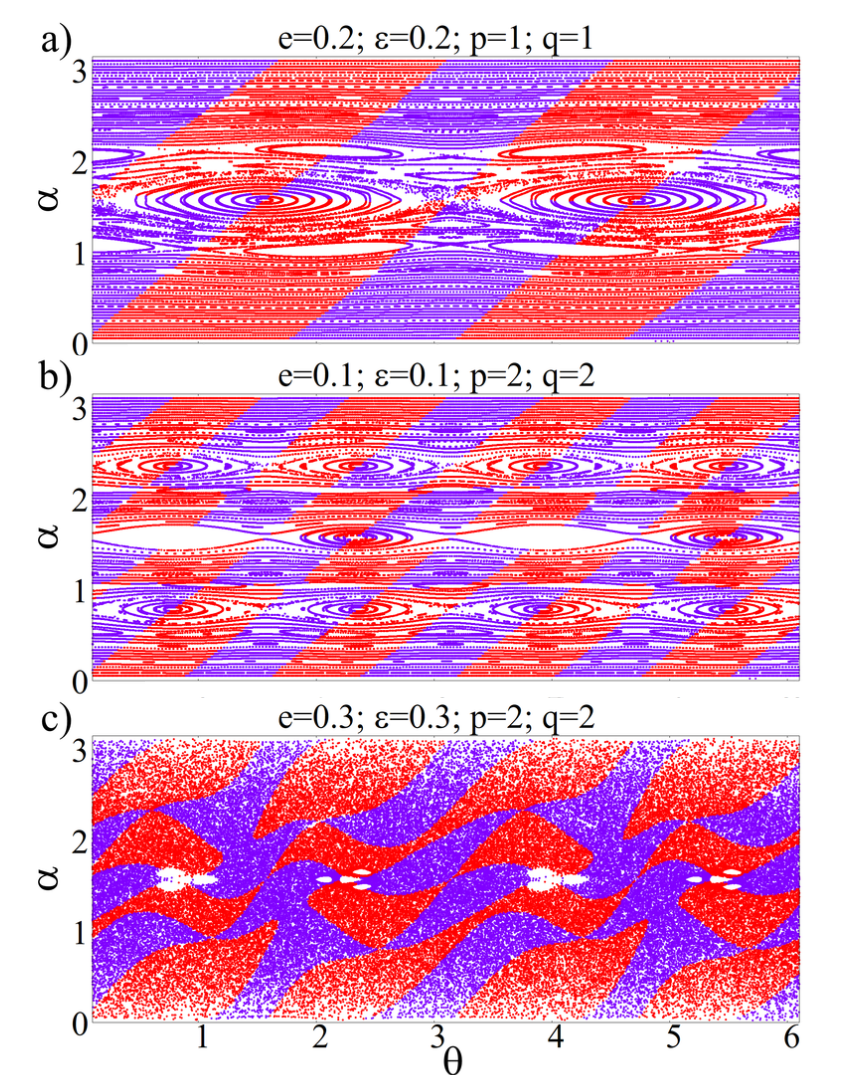}
\caption{\label{Fig5} Phase spaces and determinants of the Jacobian for the static elliptic-oval billiard with different values of $e=\epsilon$ and $p=q$, showing the multiplication of the colored regions and their subsequent deformation and merging.}
\end{figure}

A similar behavior is found in the elliptical-oval case. For $p=q$ and sufficiently small values of the control parameters $\epsilon$ and $e$, the number of colored regions follows the rule $2(p+q)^2$, as shown in panels \textit{a)} and \textit{b)} of Fig.~\ref{Fig5}.

\begin{figure*}[ht]
\includegraphics[width=1.6\columnwidth]{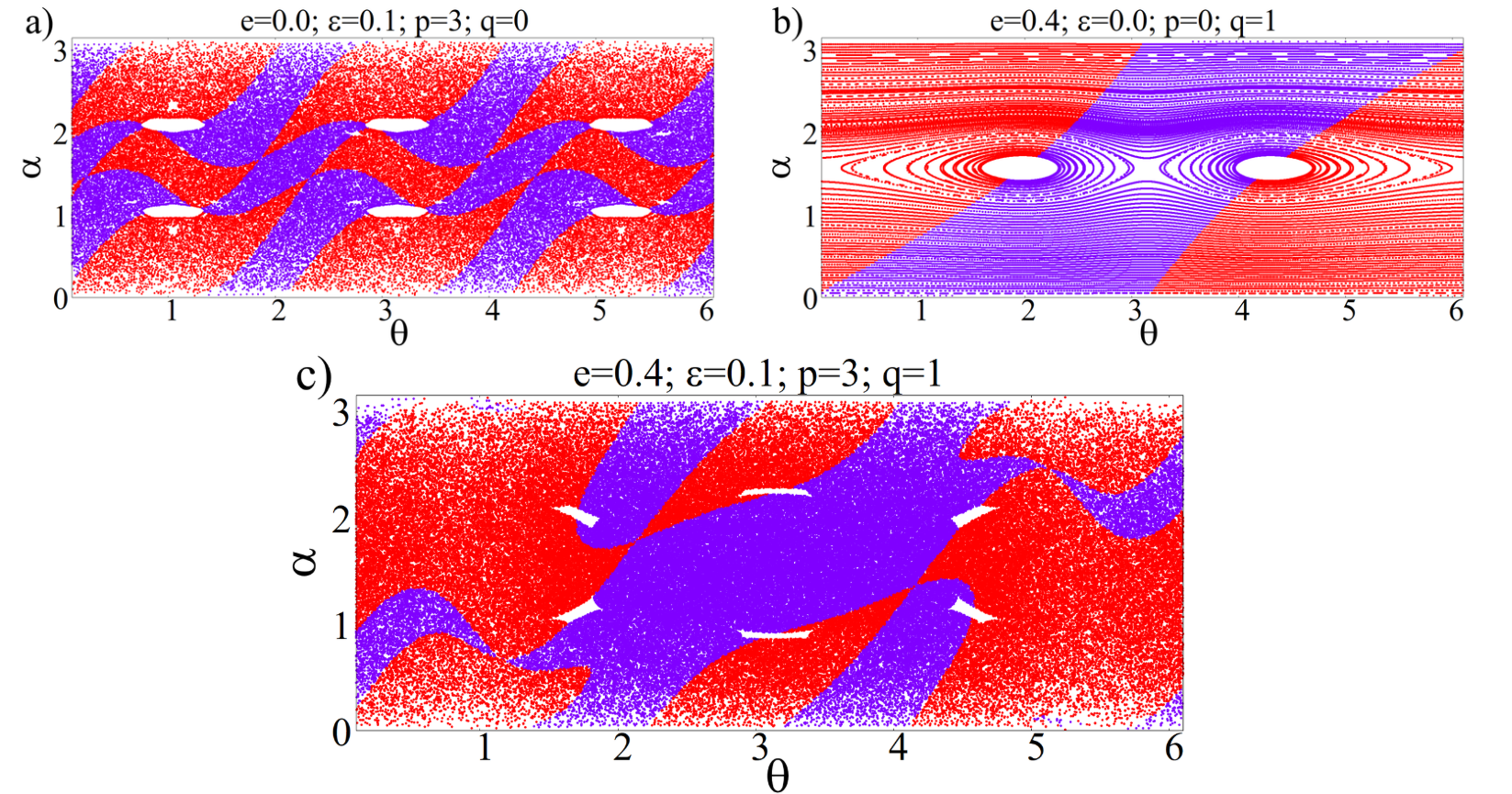}
\caption{\label{Fig6} Phase spaces and determinants of the Jacobian for the static \textit{a)}elliptic billiard; \textit{b)}oval billiard and \textit{c)} elliptical-oval billiard, overlaping the structures observed on both previous panels.}
\end{figure*}

As $\epsilon$ and $e$ increase, their contributions to the boundary geometry no longer remain equivalent (see Eq.~(\ref{front})). Consequently, the colored domains can merge and develop more irregular shapes. A similar situation occurs for $p\neq q$, where it becomes useful to interpret the resulting phase space as the superposition of the strictly oval and elliptical cases, as illustrated in Fig.~\ref{Fig6}.

The oval and elliptical phase spaces are presented in panels \textit{a)} and \textit{b)}, respectively. Fig.~\ref{Fig6} \textit{c)} exhibits structures that can be traced back to both geometries, emphasizing that the elliptical-oval billiard should not be interpreted merely as the sum of its individual contributions suggested by Eq.~(\ref{front}), but as the outcome of their geometric interaction.

\begin{figure}[h]
\includegraphics[width=\columnwidth]{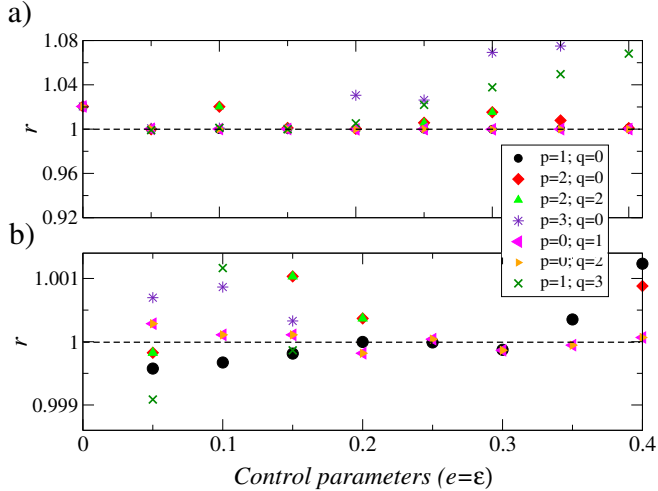}
\caption{\label{Fig7} Values of $r$ for different values of the control parameters $p$, $q$ and $\epsilon=e$. Panel \textit{b)} shows an ampliation of panel \textit{a)}, with most of the results lying within an error of 0.1\% of the expected value $r=1$.}
\end{figure}

Finally, we return to the apparent contradiction between local values $\det J \neq 1$ in $(\theta,\alpha)$ and the conservative character of the static billiard dynamics. While $\det J$ quantifies local deformation in these coordinates, global area preservation implies a compensation between contraction and expansion across phase space. To quantify this balance numerically, we introduce the ratio $r$ between the number of states in the contracting (blue, $\det J<1$) and expanding (red, $\det J>1$) regions. The values of $r$ for different control parameters are shown in Fig.~\ref{Fig7}. For convenience, we present results for $\epsilon=e$, varying $p$ and $q$, thereby encompassing the elliptical, oval and elliptical-oval cases. Most values remain very close to unity, indicating that the areas associated with $\det J>1$ and $\det J<1$ are globally balanced, as required for a conservative system.

It is worth emphasizing that states with $\det J = 1$ are observed throughout phase space. In the next section, we analyze their locations and investigate which conditions in Eq.~(\ref{detJ}) give rise to these values, together with their associated trajectories and dynamical characteristics.

\section{\label{Sec4} Fixed points and the unitary Jacoban determinant}
\begin{figure}[h]
\includegraphics[width=\columnwidth]{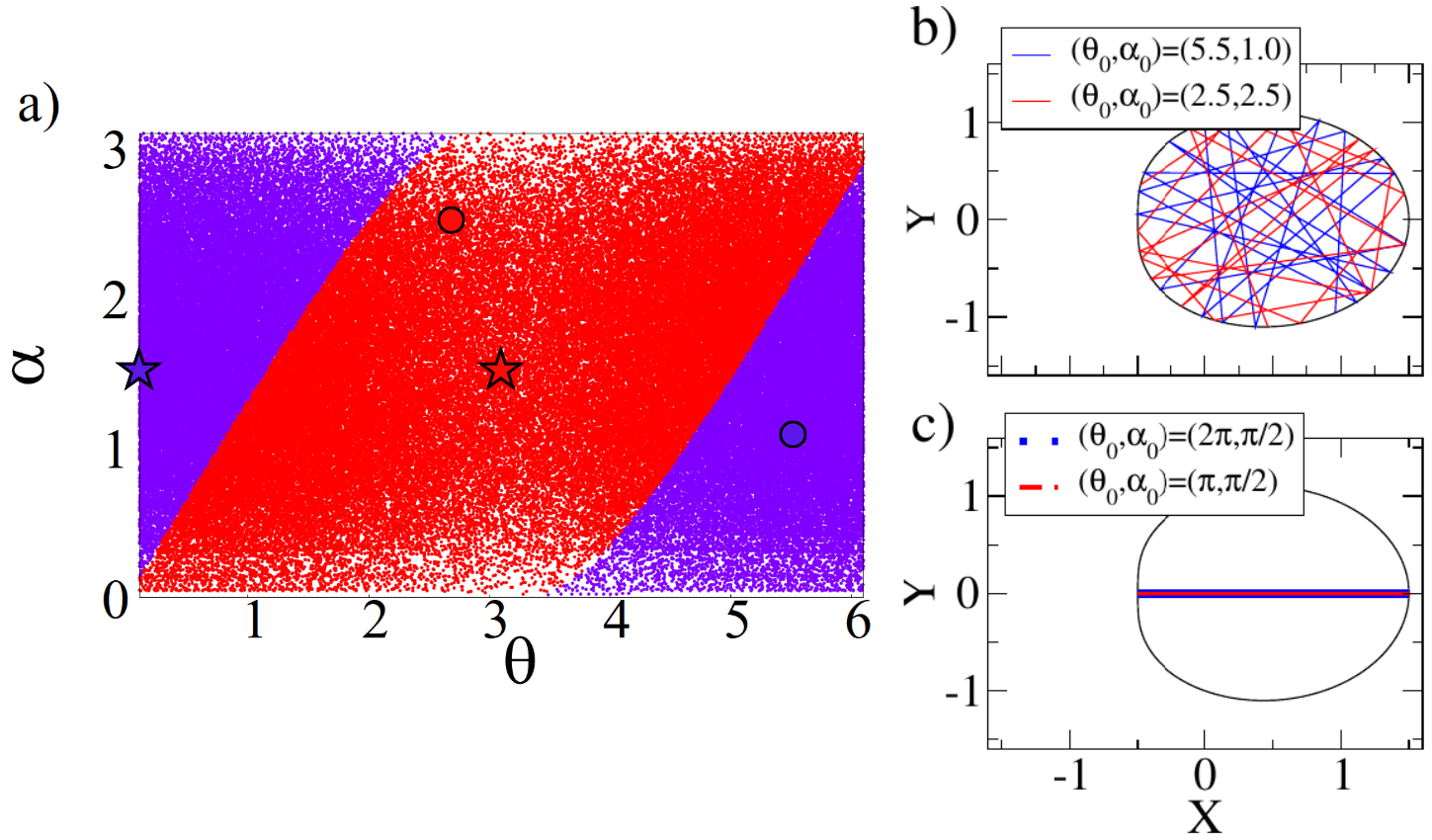}
\caption{\label{Fig8} Phase space and associated trajectories for an elliptical billiard with $\epsilon=0.5$, $e=0$, $p=1$, and $q=0$. The trajectories, for the first 20 collisions, corresponding to the circular markers in panel \textit{a)} are shown in panel \textit{b)}, whereas those associated with the star markers are shown in panel \textit{c)}.}
\end{figure}

Although the structures shown in Fig.~\ref{Fig4} may suggest distinct dynamical behaviors for the blue and red regions, trajectories initiated in these domains are not trivially distinguishable. This is illustrated in Fig.~\ref{Fig8}: the trajectories indicated by circular markers in panel \textit{a)} and shown in panel \textit{b)} do not exhibit clear qualitative differences over the first collisions. We therefore focus on the geometric structures underlying the partition, namely the boundaries between domains and their intersections. As a first indication, the trajectories associated with the star markers in Fig.~\ref{Fig8} \textit{c)}, placed near the centers of the colored regions, display stable motion over the first 20 collisions.

\begin{figure}[h]
\includegraphics[width=0.9\columnwidth]{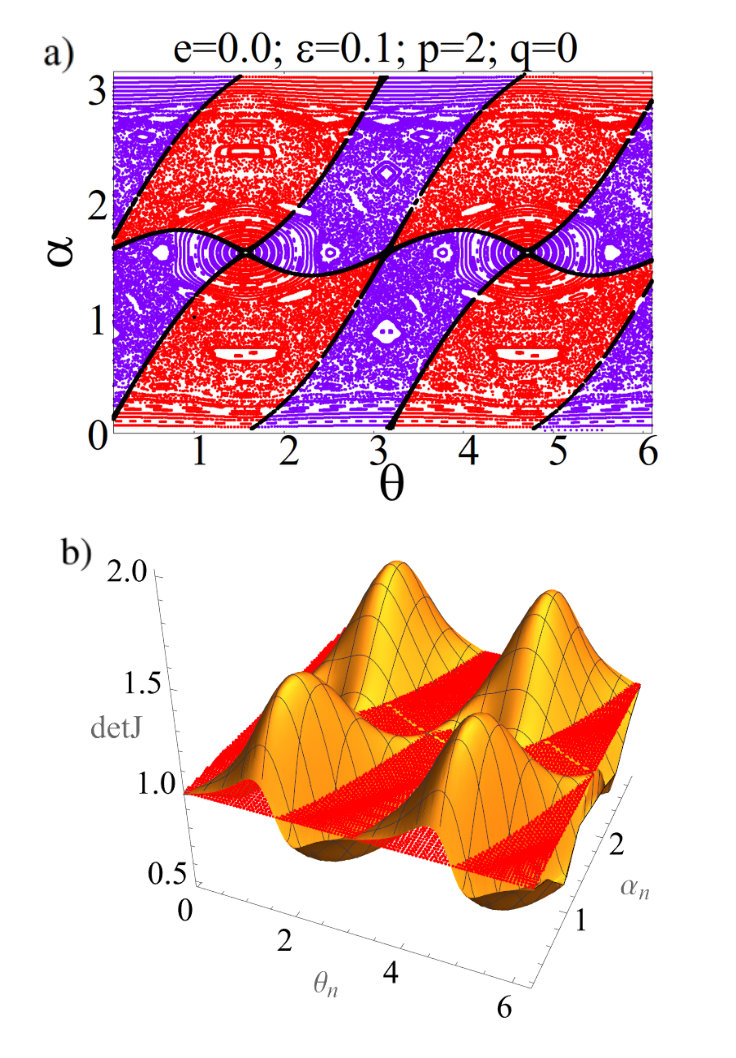}
\caption{\label{Fig9} Two and three-dimensional representations of the phase space, colored by the values of $\det J$ for an elliptical billiard. The boundary between the blue and red regions in panel \textit{a)} is given by the interection of the plane $\det J=1$ with the surface presented in panel \textit{b)}.}
\end{figure}

To further understand this organization, we characterize the boundary separating the blue and red regions, defined by the level set $\det J=1$. Panel \textit{a)} of Fig.~\ref{Fig9} shows this locus (black curves) superimposed on the colored phase space of an elliptical billiard. Panel \textit{b)} provides a three-dimensional representation: the red surface corresponds to the solutions of Eq.~(\ref{detJ}) satisfying $\det J=1$, and its intersection with the phase-space plane reproduces precisely the black curves observed in panel \textit{a)}. These curves therefore delimit domains of local contraction and expansion in the chosen coordinates. Moreover, the three-dimensional representation in Fig.~\ref{Fig9} suggests an interpretation of the ratio $r$, introduced in the previous section, in terms of the balance between contributions above and below the plane $\det J=1$.

\begin{figure*}[ht]
\hspace{-10pt}
\includegraphics[width=2.1\columnwidth]{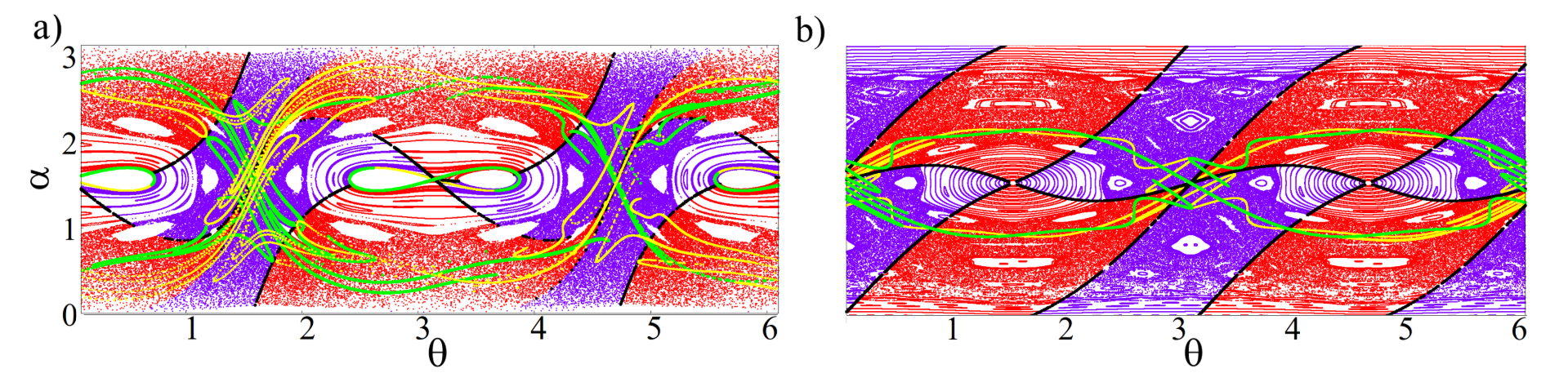}
\caption{\label{Fig10} Phase spaces colored by the values of $\det J$ for an elliptical and oval billiard, along with the stable and unstable manifolds, in green and yellow respectively. The elliptical billiard in panel \textit{a)} shows hyperbolic saddle points for all four crossing points $\left(\frac{\pi}{2},\frac{\pi}{2}\right)$, $\left(\pi,\frac{\pi}{2}\right)$ , $\left(\frac{3\pi}{2},\frac{\pi}{2}\right)$ and $\left(2\pi,\frac{\pi}{2}\right)$ with control parameters $\epsilon=0.0$, $e=0.4$, $p=0$ and $q=2$. In panel \textit{b)} the results for the oval billiard, with  $\epsilon=0.1$, $e=0.0$, $p=2$ and $q=0$ are presented. We identify two hyperbolic saddle points in $\left(\pi,\frac{\pi}{2}\right)$, $\left(2\pi,\frac{\pi}{2}\right)$ and two elliptical fixed points in $\left(\frac{\pi}{2},\frac{\pi}{2}\right)$, $\left(\frac{3\pi}{2},\frac{\pi}{2}\right)$}
\end{figure*}

Having characterized both the domains and their boundaries, we now investigate the dynamical structures revealed by intersections of the $\det J=1$ curves. Fig.~\ref{Fig10} shows stable and unstable manifolds constructed from crossing points of the black curves located at $\left(\frac{\pi}{2},\frac{\pi}{2}\right)$, $\left(\pi,\frac{\pi}{2}\right)$, $\left(\frac{3\pi}{2},\frac{\pi}{2}\right)$ and $\left(2\pi,\frac{\pi}{2}\right)$ for elliptical and oval billiards, respectively. Hyperbolic saddle points are identified in panel \textit{a)} for all four crossings with $\epsilon=0.0$, $e=0.4$, $p=0$ and $q=2$. The figure-eight structure near $\left(\frac{\pi}{2},\pi\right)$ and $\left(2\pi,\frac{\pi}{2}\right)$ highlights a homoclinic connection, where stable and unstable manifolds intersect. Although the crossings appear for $e \neq 0$, the corresponding structure becomes clearly identifiable only for $e \gtrsim 0.4$, indicating a change in stability of fixed points that were previously elliptic. Geometrically, increasing $\epsilon$ and $e$ enhances boundary deformation, which in turn affects the stability of specific periodic orbits. In panel \textit{b)}, corresponding to the oval billiard with $\epsilon=0.1$, $e=0.0$, $p=2$ and $q=0$, fixed points are also identified: two hyperbolic saddles at $\left(\pi,\frac{\pi}{2}\right)$ and $\left(2\pi,\frac{\pi}{2}\right)$, and two elliptic fixed points at $\left(\frac{\pi}{2},\frac{\pi}{2}\right)$ and $\left(\frac{3\pi}{2},\frac{\pi}{2}\right)$. In all cases, the fixed points coincide with intersections of the $\det J = 1$ curves.

\begin{figure}[h]
\includegraphics[width=0.7\columnwidth]{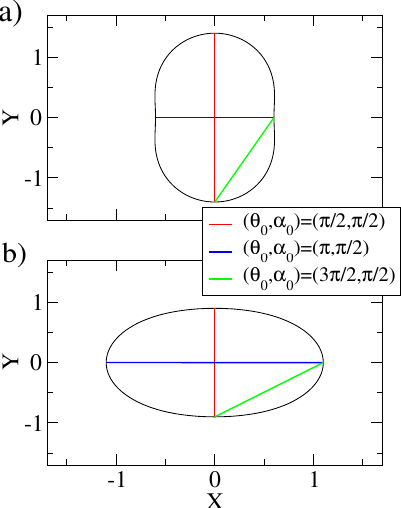}
\caption{\label{Fig11} Trajectories associated with the fixed points shown in Fig.~\ref{Fig10} for the control parameters: \textit{a)} $\epsilon = 0.0$, $e = 0.4$, $p = 0$, and $q = 2$; and \textit{b)} $\epsilon = 0.1$, $e = 0.0$, $p = 2$, and $q = 0$ with initial conditions presented in the figure. The first 20 collisions indicate that the trajectories remain stable.}
\end{figure}

These results indicate that the level set $\det J=1$ provides a deformation-based skeleton of phase space that is strongly correlated with the classical manifold description. The connection between both approaches lies in the unstable fixed points, which are intersected by both the invariant manifolds and the $\det J = 1$ curves. The trajectories associated with the fixed points shown in Fig.~\ref{Fig10} lie along symmetry axes of the billiard: (i) along the vertical and horizontal axes for $\left(\frac{\pi}{2}, \frac{\pi}{2}\right)$ and $\left(\pi, \frac{\pi}{2}\right)$, respectively, and (ii) connecting extremal points of these axes for $\left(\frac{3\pi}{2},\frac{\pi}{2}\right)$. An initial condition placed exactly at any of these points generates a trajectory that remains on the same path indefinitely. Fig.~\ref{Fig11} presents the first 20 collisions starting from each fixed point for the same boundaries shown in Fig.~\ref{Fig10}.

\begin{figure}[h]
\hspace{-10pt}
\includegraphics[width=\columnwidth]{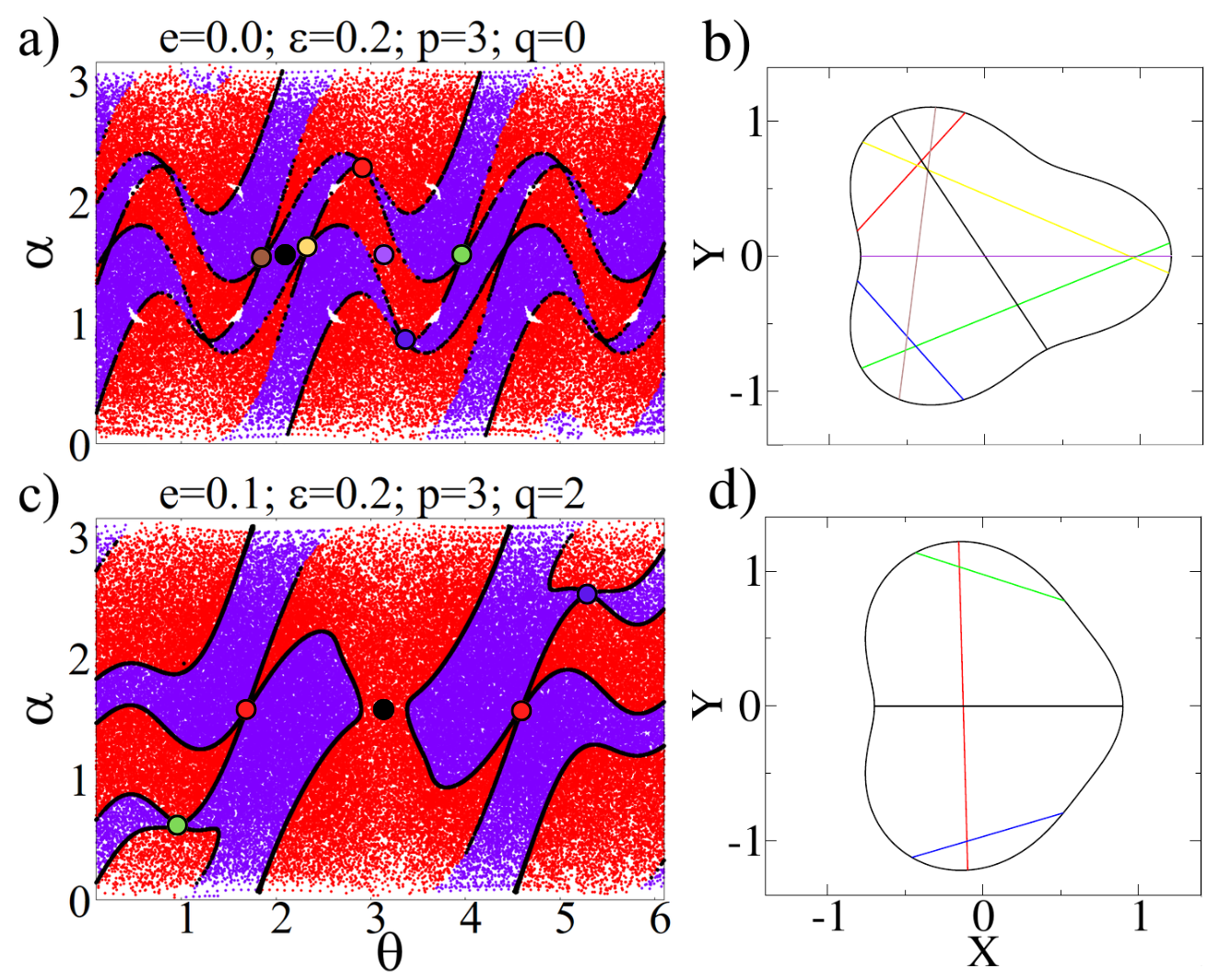}
\caption{\label{Fig12} Phase space colored by the values of $\det J$ for different control parameters. Panels \textit{b)} and \textit{d)} show sketches of the trajectories associated with the initial conditions indicated by the respective colored circles in panels \textit{a)} and \textit{c)}.}
\end{figure}

In Fig.~\ref{Fig12} we extend this analysis to other geometries, including the elliptical-oval billiard. Sketches of trajectories located at the centers of colored regions and at boundary crossings allow identification of stable orbits distributed along the boundary. The results for $e=0.0$, $\epsilon=0.2$, $p=3$ and $q=0$ in panels \textit{a)} and \textit{b)} of Fig.~\ref{Fig12} reveal not only the expected stable trajectory at $\left(\pi,\frac{\pi}{2}\right)$, but several others connecting the lobes of the billiard. Thus, increasing $p$ and $q$ leads not only to a proliferation of colored structures, but also of associated fixed points. Variations in $\epsilon$ and $e$ deform these regions and may alter the position and stability of the corresponding periodic orbits. Panels \textit{c)-d-)} of Fig.~\ref{Fig12} and Figs.~\ref{Fig5}-\ref{Fig6} illustrate the potential of this approach in revealing nontrivial structures in complex billiard geometries. As previously discussed, these crossings are not simply the superposition of those from strictly oval and elliptical cases, but emerge from their interplay. This deformation-based perspective thus provides a complementary tool for uncovering structures within static billiard dynamics.

This approach becomes particularly valuable in identifying fixed points, whose determination directly from Eq.~(\ref{map}) is not straightforward. The first nontrivial fixed points are of period two, since a particle requires at least two iterations to return to the same geometrical configuration. Substituting $\theta_{n+1}=\theta_n=\theta^*$ and $\alpha_{n+1}=\alpha_n=\alpha^*$ into Eq.~(\ref{map}) yields $\alpha^*=\frac{\pi}{2}$ and $h(\theta^*)=0$, which provides limited information due to the transcendental character of $h(\theta)$. We therefore adopt an alternative approach based on the Jacobian determinant. Assuming that for a period-two fixed point $F(P)=Q\neq P$ and $F(Q)=P$, namely $P\rightarrow Q \rightarrow P$, the determinant satisfies \cite{strog}
\begin{equation}
\det J_{F^2}(P)=\det J_F(Q)\cdot \det J_F(P).
\end{equation}
For Eq.~(\ref{detJ}) this leads to (see the Appendix for details)
\begin{equation}
\det J_{F^2}=\left(-\frac{X'(\theta^*)}{X'(\theta_{n+1})}\right)\left(-\frac{X'(\theta_{n+1})}{X'(\theta^*)}\right)=1.
\end{equation}
Thus, the composed map $F^2$, which fixes the period-two orbit, has unit determinant at the fixed point. This property does not necessarily extend to higher-order periodic orbits, as shown in the Appendix. For period $n>2$, even at the fixed point, the determinant is modulated by the angular variables $\alpha$ and $\phi$, and $\vert\det J_{F^n}\vert = 1$ holds only under specific conditions. Fixed points are therefore not restricted to the curve defined by $\det J=1$. However, those lying on this curve satisfy the stronger property that $\det J_F$ equals unity at each step along the orbit, i.e., $\det J_F(Q)=\det J_F(P)=....=1$. Other fixed points may not satisfy this condition at every iteration, but reversibility ensures a global balance between local expansion and contraction along the periodic trajectory.

\section{\label{Sec5} Summary and conclusions}

In this work we introduced a deformation-based framework for analyzing the phase space of static billiard systems through the Jacobian determinant computed in noncanonical coordinates. Although the angular variables $(\theta,\alpha)$ do not form a canonical pair and therefore $\det J$ is not identically equal to unity, the system remains conservative. The determinant thus encodes local geometric deformation rather than dissipation.

We showed that $\det J$ partitions phase space into organized domains associated with local expansion ($\det J>1$) and contraction ($\det J<1$). The number, arrangement, and deformation of these domains are governed by the boundary parameters $\epsilon$, $e$, $p$, and $q$. In particular, the elliptical–oval billiard demonstrates that these structures emerge from a nonlinear interplay between purely elliptical and purely oval geometries, rather than from a simple superposition of their individual contributions.

Despite local variations of $\det J$, numerical analysis confirms that the global balance between expanding and contracting regions is preserved over a wide range of parameters. The near-unity ratio between these regions is therefore consistent with global area preservation when the dynamics is expressed in noncanonical variables, providing a geometric consistency check of conservativity.

A central structural role is played by the curves defined by $\det J=1$. These curves act as deformation boundaries that intersect unstable fixed points and correlate with invariant manifolds. In particular, for period-two orbits we proved analytically that the composed map $F^2$ satisfies $\det J_{F^2}=1$, establishing exact unit determinant at the periodic cycle. For higher-period orbits, angular contributions modulate the determinant, yet reversibility guarantees a global compensation between local expansion and contraction along each complete orbit.

Taken together, these results indicate that the Jacobian determinant, even when computed in noncanonical variables, carries nontrivial structural information about conservative billiard dynamics. Rather than contradicting area preservation, local variations of $\det J$ reveal a geometric organization that complements the classical manifold-based description. This deformation-based perspective may be extended to other billiard geometries and, more broadly, to conservative dynamical systems formulated in noncanonical coordinates, where geometric deformation and conservation coexist within a coherent dynamical structure.

\begin{acknowledgments}
The authors gratefully acknowledge Prof. Dr. Everton Santos Medeiros and Prof. Dr. André Luis Prando Livorati for valuable discussions and insightful comments.
\end{acknowledgments}

\section*{Data Availability Statement}

The data that support the findings of this study are available from the corresponding author upon reasonable request.

\subsection*{CRediT authorship contribution statement}

\textbf{Anne Kétri P. da Fonseca}: Software, Validation, Formal analysis, Data curation, Writing – Original Draft, Writing – Review \& Editing. 
\textbf{André L. P. Livorati}: Conceptualization, Methodology.
\textbf{Rene O. Medrano-T}: Methodology, Supervision, Investigation, Writing – Review \& Editing. 
\textbf{Diego F. M. Oliveira}: Conceptualization, Investigation, Writing – Review \& Editing. 
\textbf{Edson D. Leonel}: Conceptualization, Methodology, Supervision, Writing – Review \& Editing. 

\subsection*{Funding}

Anne Kétri P. da Fonseca received funding from CAPES under Grant Agreement No.~$88887.990665/2024-00$. 
Edson D. Leonel received funding from Brazilian agencies CNPq under Grant Agreements No.~$301318/2019-0$ and $304398/2023-3$, and from FAPESP under Grant Agreements No.~$2019/14038-6$ and No.~$2021/09519-5$.

\subsection*{Conflicts of interest}

The authors declare the following financial interests and relationships that could be considered potential competing interests: Edson D. Leonel reports that equipment and supplies were provided by the São Paulo State University, Institute of Geosciences and Exact Sciences. He also reports a relationship with the State of São Paulo Research Foundation involving board membership. Anne Kétri P. da Fonseca, Rene O. Medrano-Torricos and Diego F. M. Oliveira declare no conflicts of interest relevant to the content of this article.
\appendix*

\section{Derivation of $\det J_{F^2}=1$}

For a period-two orbit of a mapping $F$ satisfying $F(P)=Q$ and $F(Q)=P$, the determinant of the composed map satisfies \cite{strog}
\begin{equation}
\det J_{F^2}(P)=\det J_F(Q)\cdot \det J_F(P).
\end{equation}

Let $P=(\theta^*,\alpha^*)$ and $Q=F(P)=(\theta_{n+1},\alpha_{n+1})$. Using Eq.~(\ref{detJ}) together with the relation $Y'(\theta)=X'(\theta)\tan\phi$, each step contributes a geometric factor of the form
\begin{equation}
\det J_k = -\frac{X'(\theta_k)}{X'(\theta_{k+1})}.
\end{equation}

Therefore, for the period-two orbit,
\begin{equation}
\det J_{F^2}
=\left(-\frac{X'(\theta^*)}{X'(\theta_{n+1})}\right)
\left(-\frac{X'(\theta_{n+1})}{X'(\theta^*)}\right)
=1.
\end{equation}

The exact cancellation reflects the telescopic structure of the geometric factors and establishes unit determinant for the composed map along a period-two orbit.

For higher-period orbits $P_1 \rightarrow P_2 \rightarrow \cdots \rightarrow P_n \rightarrow P_1$, the determinant of the composed map is
\begin{gather*}
\det J_{F^n} = \prod_{k=1}^n \det J_k \\= \prod_{k=1}^n \left(-\frac{X'(\theta_k)[\tan\phi_k - \tan(\phi_k + \alpha_k)]}{X'(\theta_{k+1})[\tan\phi_{k+1} - \tan(\alpha_k + \phi_{k+1})]}\right)
\end{gather*}
or simply 
\begin{equation}
\det J_{F^n} = (-1)^n \cdot \prod_{k=1}^n\frac{ [\tan\phi_k - \tan(\phi_k + \alpha_k)]}{[\tan\phi_{k+1} - \tan(\alpha_k + \phi_{k+1})]}
\end{equation}
As in the period-two case, the geometric factors involving $X'(\theta_k)$ cancel telescopically. However, the angular contributions generally do not cancel completely. Consequently, $\vert \det J_{F^n}\vert = 1$ holds only under specific geometric constraints along the orbit.

\nocite{*}
\bibliography{aipsamp}

@PREAMBLE{
 "\providecommand{\noopsort}[1]{}" 
 # "\providecommand{\singleletter}[1]{#1}%" 
}

@BOOK{briefhistory,
    author      =   "Predrag Cvitanovic",
    title       =   "A Brief History of Chaos",
    year        =   "1989",
    publisher   =   "Georgia Institute of Technology",
    address     =   " Atlanta, Georgia - Estados Unidos da América "
}

@BOOK{Arnold,
  author    = {Arnold, Vladimir I.},
  title     = {Mathematical Methods of Classical Mechanics},
  publisher = {Springer},
  year      = {1989},
  edition   = {2},
  series    = {Graduate Texts in Mathematics},
  volume    = {60},
  address   = {New York},
  isbn      = {978-0-387-96890-2}
}

@BOOK{lichtenberg,
    author      =   "Lichtenberg, AJ and Lieberman, MA",
    title       =   "Regular and Chaotic Dynamics",
    year        =   "1992",
    publisher   =   "Springer",
    address     =   "Verlag, Nova Iorque - Estados Unidos da América"
}

@BOOK{chernov,
  title        = "Chaotic Billiards",
  author       = "Chernov, Nikolai and Markarian, Roberto",
  publisher    = "American Mathematical Society",
  year         = "2006",
  address      = "Providence, RI, USA"
}

@INBOOK{livrodenisspringer,
author="Leonel, Edson Denis",
title="Time Dependent Billiards",
bookTitle="Scaling Laws in Dynamical Systems",
year="2021",
publisher="Springer Singapore",
address="Singapore",
pages="181--190",
doi="10.1007/978-981-16-3544-1_13"
}

@ARTICLE{berry,
    author      =   "Berry, MV",
    title       =   "Regularity and chaos in classical mechanics, illustrated by three deformations of a circular 'billiard'
",
    journal     =   "European Journal of Physics",
    volume      =   "2",
    number      =   "2",
    pages       =   "91--102",
    year        =   "1981",
    DOI         =   "10.1088/0143-0807/2/2/006"
}

@ARTICLE{sinai,
    author      =   "Yakov Sinai",
    title       =   "Dynamical systems with elastic reflections. Ergodic properties of dispersing billiards",
    journal     =   "Russian Mathematical Survey",
    volume      =   " 25",
    number      =   " 2",
    pages       =   "137 -- 189",
    year        =   "1970",
    DOI         =   "10.1070/rm1970v025n02abeh003794"
}

@ARTICLE{ref1,
title = {Statistical properties of a dissipative kicked system: Critical exponents and scaling invariance},
journal = {Physics Letters A},
volume = {376},
number = {5},
pages = {723-728},
year = {2012},
doi = {https://doi.org/10.1016/j.physleta.2011.12.031},
author = {Diego F.M. Oliveira and Marko Robnik and Edson D. Leonel}
}

@article{livoratifaixas,
  title = {Global ballistic acceleration in a bouncing-ball model},
  author = {Kroetz, Tiago and Livorati, Andr\'e L. P. and Leonel, Edson D. and Caldas, Iber\^e L.},
  journal = {Phys. Rev. E},
  volume = {92},
  issue = {1},
  pages = {012905},
  numpages = {11},
  year = {2015},
  month = {Jul},
  publisher = {American Physical Society},
  doi = {10.1103/PhysRevE.92.012905},
  url = {https://link.aps.org/doi/10.1103/PhysRevE.92.012905}
}

@book{strog,
    author      =   "Strogatz, S.H. ",
    title       =   "Nonlinear Dynamics and Chaos: With Applications to Physics, Biology, Chemistry, and Engineering",
    year        =   "2015",
    publisher   =   "CRC Press",
    address     =   "Florida, Estados Unidos da América"
}

@article{ref3,
title = {A symmetry break in energy distribution and a biased random walk behavior causing unlimited diffusion in a two dimensional mapping},
journal = {Physica A: Statistical Mechanics and its Applications},
volume = {436},
pages = {909-915},
year = {2015},
doi = {https://doi.org/10.1016/j.physa.2015.05.065},
author = {Diego F.M. Oliveira and Mario R. Silva and Edson D. Leonel}
}

@article{ref4, 
    author      =   "Bunimovich, LA",
    title       =   "On the ergodic properties of nowhere dispersing billiards",
    journal     =   "Communications in Mathematical Physics",
    volume      =   "65",
    number      =   "3",
    pages       =   "295-312",
    year        =   "1979",
    DOI         =   "10.1007/BF01197884"
}

@ARTICLE{onthe,
   author       = "Diego F.M. Oliveira and Edson D. Leonel",
   title        = "On the dynamical properties of an elliptical–oval billiard with
static boundary",
   journal      = "Commun Nonlinear Sci Numer Simulat",
   volume       = "15", 
   pages        = "1092-1102",
   year         = "2010"}

@article{critic,
title = {Suppressing Fermi acceleration in a two-dimensional non-integrable time-dependent oval-shaped billiard with inelastic collisions},
journal = {Physica A: Statistical Mechanics and its Applications},
volume = {389},
number = {5},
pages = {1009-1020},
year = {2010},
doi = {https://doi.org/10.1016/j.physa.2009.10.036},
author = {Diego F.M. Oliveira and Edson D. Leonel}
}

\end{document}